\documentclass[twocolumn,superscriptaddress,pr]{revtex4-1}
\usepackage{amsmath,amssymb}
\usepackage{graphicx}
\usepackage{color}
\usepackage{times}
\usepackage[colorlinks,bookmarks=false,citecolor=blue,linkcolor=red,
urlcolor=blue]{hyperref}

\begin{document}

\author{F.~Kolley}
\affiliation{Department of Physics and Arnold Sommerfeld 
Center for Theoretical Physics, Ludwig-Maximilians-Universit\"at 
M\"unchen, D-80333 M\"unchen, Germany}
\author{S.~Depenbrock}
\affiliation{Department of Physics and Arnold Sommerfeld 
Center for Theoretical Physics, Ludwig-Maximilians-Universit\"at 
M\"unchen, D-80333 M\"unchen, Germany}
\author{I.~P.~McCulloch}
\affiliation{School of Physical Sciences, The University of Queensland, Brisbane, 
QLD 4072, Australia}
\author{U.~Schollw\"ock}
\affiliation{Department of Physics and Arnold Sommerfeld 
Center for Theoretical Physics, Ludwig-Maximilians-Universit\"at 
M\"unchen, D-80333 M\"unchen, Germany}
\author{V.~Alba}
\affiliation{Department of Physics and Arnold Sommerfeld 
Center for Theoretical Physics, Ludwig-Maximilians-Universit\"at 
M\"unchen, D-80333 M\"unchen, Germany}

\date{\today}

\title{Entanglement spectroscopy of SU(2)-broken phases   
in two dimensions}

\begin{abstract} 

In magnetically ordered systems the breaking of $SU(2)$ symmetry in the thermodynamic 
limit is associated with the appearance of a special type of low-lying excitations in 
finite size energy spectra, the so called tower-of-states (TOS). In the present work
we numerically demonstrate that there is a correspondence between the $SU(2)$ tower of states 
and the lower part of the {\it ground state} entanglement spectrum (ES). Using state-of-the-art 
DMRG calculations, we examine the ES of the 2D antiferromagnetic $J_1$-$J_2$ Heisenberg 
model on both the triangular and kagom\'e lattice. At large ferromagnetic $J_2$ the model 
exhibits a magnetically ordered ground state. Correspondingly, its ES contains a family 
of low-lying levels that are reminiscent of the energy tower of states. Their behavior 
(level counting, finite size scaling in the thermodynamic limit) sharply reflects TOS 
features, and is characterized in terms of an effective entanglement Hamiltonian that 
we provide. At large system sizes TOS levels are divided from the rest by an entanglement 
gap. Our analysis suggests that (TOS) entanglement spectroscopy provides an alternative 
tool for detecting and characterizing $SU(2)$-broken phases using DMRG.

\end{abstract}


\maketitle



\section{Introduction}

Recent years have witnessed an increasing interest in  entanglement related 
quantities (and quantum information concepts in general) as new tools  to 
understand the behavior of quantum many body systems~\cite{amico-2008}. 
Very recently the entanglement spectrum~\cite{li-2008} (ES) has established 
itself as a new prominent research topic. Considering the bipartition of a system 
into parts $A$ and $B$, the ES, $\{\xi_i\}$, is constructed from the 
Schmidt decomposition
\begin{equation}  \label{eq:schmidt_ES_defn}
|\psi\rangle=\sum_i e^{-\xi_i/2}|\psi_i^A\rangle
\otimes |\psi_i^B\rangle. 
\end{equation}
Here $|\psi\rangle$ is the ground state, and the states $|\psi_i^A\rangle$ 
($|\psi_i^B\rangle$) provide an orthonormal basis for subsystem $A$ ($B$).  
The ES $\{\xi_i\}$ can also be interpreted as the spectrum 
of the so called entanglement Hamiltonian ${\mathcal H}_E\equiv -\log\rho_A$, 
where the reduced density matrix $\rho_A$ is obtained by tracing out
part $B$ in the full system density matrix $|\psi\rangle\langle\psi|$.

While in one dimensional (1D) systems the structure of ES is related to 
integrability~\cite{peschel-2009,alba-2011,lepori-2013}  and (for gapless systems) 
to conformal invariance~\cite{calabrese-2008,dechiara-2012,lepori-2013,
laeuchli-2013}, higher dimensions are by far less explored. In particular, 
most of the recent literature on two dimensional (2D) systems focused on 
ES properties in topological phases~\cite{li-2008,topological}. 

In more standard (i.e. non topological)  2D systems, although 
some results are available~\cite{cirac-2011,tanaka-2012,lou-2013,james-2013}, 
much less is known. Nevertheless, it has 
been established recently that in  systems displaying ordered ground states 
(in the thermodynamic limit), with breaking of a {\it continuous} symmetry,  
the lower part of the ES is in correspondence with the so called ``tower of 
states'' (TOS) spectrum~\cite{met-grov-2011,alba-2012}. This describes the low 
energy structure of {\it finite size} spectra in systems  that spontaneously 
break a continuous symmetry.  In combination with exact diagonalization 
techniques, tower of states spectroscopy is routinely used to detect symmetry 
broken phases~\cite{bernu-1992,bernu-1994,lech-1995,lech-1997,laeuchli-2005,
shannon-2006,sindzingre-2009,penc-2011}.

So far tower of states structures in ES have only been observed numerically 
in the superfluid phase of the 2D Bose-Hubbard model~\cite{alba-2012}, 
where the formation of a Bose condensate is associated with the breaking 
of a $U(1)$ gauge symmetry (reflecting conservation of the total number 
of particles in finite systems). The resulting TOS spectrum, however,  
(and the lower part of the ES thereof) is ``trivial'' with one level 
(excitation) per particle number sector~\cite{alba-2012}. 

Richer behavior is expected in $SU(2)$-broken phases, where different 
$SU(2)$ breaking patterns (i.e. N\'eel states) give rise to different 
structures in the energy TOS. For instance, for N\'eel order 
with more than two ferromagnetic sublattices (associated with full breaking 
of $SU(2)$) the spin resolved TOS spectrum exhibits a family of levels 
(i.e. more than one level) in each spin sector~\cite{lhuillier_arxiv2005}. 

In this Article we demonstrate that this richer structure is reflected 
in the lower part of the ES, providing a more stringent check of the 
correspondence between tower of states and entanglement spectrum. To be 
specific, we focus on the 2D  Heisenberg model with nearest and next-nearest 
neighbor interaction ($J_1$ and $J_2$ respectively). We consider both the 
kagom\'e (KHA) and the triangular lattice (THA), restricting ourselves to 
ferromagnetic $J_2$ ($J_2=-1$), to ensure a magnetically ordered ground state 
on both lattices. In order to take advantage of the $SU(2)$ invariance of the 
model we employ non-abelian ($SU(2)$-symmetric) DMRG simulations. 

Our results are summarized as follows. In both the $J_1$-$J_2$ KHA and 
THA, in the symmetry broken phase, the lower part of the ES (resolved with 
respect to the block spin $S_A$) sharply reflects the same TOS structure as 
the physical bulk Hamiltonian. Low-lying ES levels are organized 
into families, each corresponding to a different $S_A$ and containing 
more than one level (in contrast to the Bose-Hubbard model where the TOS 
structure is ``trivial''). The counting of TOS levels in each $S_A$ sector 
reflects the corresponding counting in the energy TOS. 

The TOS-like structure  is divided from  higher levels by an entanglement 
gap, which remains finite (or vanishes logarithmically) in the thermodynamic 
limit (as found in the Bose-Hubbard ES~\cite{alba-2012}). All ES levels below 
the gap are  degenerate in the thermodynamic limit,  and their finite size 
behavior is fully understood within the framework of the TOS-ES correspondence. 
Oppositely, for finite systems, ES levels within each TOS family are not 
exactly degenerate (similarly to energy TOS~\cite{lhuillier_arxiv2005}) giving rise 
to intriguing entanglement (TOS) substructures. The main features of TOS levels 
(TOS substructures, finite size behaviors) are quantitatively characterized by an 
approximate mapping between the entanglement Hamiltonian and the physical bulk 
Hamiltonian. 

Finally, as an additional point, we investigate  the effect of boundary conditions 
on the TOS structure. To this purpose we  consider the ES of the $J_1$-$J_2$ 
KHA ($J_2/J_1=-1$) on the torus geometry, which  has the net effect of introducing 
two boundaries (edges) between subsystem $A$ and $B$. We find that in $SU(2)$ broken 
phases the structure of the ES is weakly affected  by boundaries, reflecting the 
bulk origin of TOS excitations. This is dramatically different in gapped 
phases~\cite{alba-2011,alba-2012} or in FQH systems~\cite{topological} where 
the ES obtained from bipartitions with multiple edges can be  constructed  
combining  single-edge ES. 

On the methodological side, our analysis suggests that entanglement TOS 
spectroscopy, combined with $SU(2)$ symmetric DMRG, could provide a potentially 
powerful tool to detect and characterize magnetically ordered ground states. 
Also, while conventional energy TOS spectroscopy requires the calculation of 
several excited states (which is computationally expensive in DMRG),  ES are 
readily obtained from {\it ground state} wavefunctions only.

The Article is organized as follows. Section~\ref{models} introduces the 
$J_1$-$J_2$ Heisenberg model on both the kagom\'e and triangular lattice. 
Some basic facts about conventional (energy) tower of states  spectroscopy in
$SU(2)$ broken phases are given in section~\ref{energy_tos}. In 
section~\ref{tos_es} we establish the correspondence between the tower of states 
and the low-lying part of the ES (cf.~\ref{TOS-ES}). This is numerically verified  
in~\ref{kagome-ES} for the $J_1$-$J_2$ kagom\'e and triangular lattice 
Heisenberg model. The fine structure (i.e. entanglement substructures) of ES 
levels building the TOS is detailed in~\ref{asy_top}. Finite size 
behavior of the (TOS) ES and its dependence on boundary conditions are discussed 
in section~\ref{fin_size}. Section~\ref{conclusions} concludes the Article.

\section{Models and method}
\label{models}

\begin{figure}[t]
\includegraphics[width=1\columnwidth]{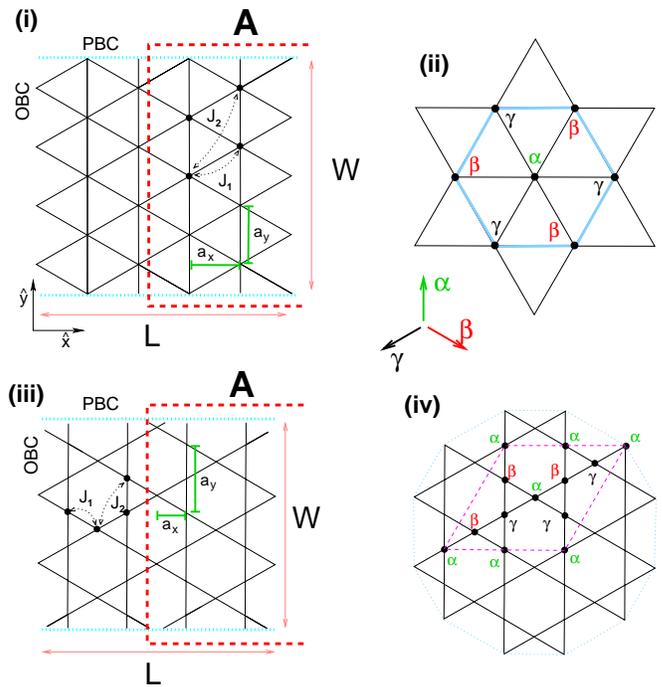}
\caption{ The $J_1$-$J_2$ Heisenberg model on the triangular (THA) 
 and kagom\'e (KHA) lattice. (i) Example of triangular cylinder 
 of length $L$ and width $W$ (measured respectively in units 
 of $a_x$ and $a_y$). Total number of sites is given as $W\times
 L$. Spins are at the vertices of the lattice. Periodic boundary 
 conditions are used along the vertical direction. $J_1$ ($J_2$) 
 is the interaction strength  between nearest (next-nearest) 
 neighbor spins. In this work we restrict ourselves to antiferromagnetic 
 (ferromagnetic) $J_1$ ($J_2$) (i.e. $J_1>0$, $J_2<0$). The dashed 
 line is to illustrate the bipartition into two subsystems.
 (ii) Ordering pattern of the THA.  Three  possible orientations 
 of the sublattice spins are denoted as $\alpha,\beta,\gamma$. 
 The angle formed by any pair of spins   is $2\pi/3$. (iii) 
 Heisenberg $J_1$-$J_2$ on the kagom\'e lattice (KHA). 
 Total number of spins is now $3\times W\times L$. (iv) 
 Ordering pattern of the $J_1$-$J_2$ KHA ($\sqrt{3}\times\sqrt{3}$ 
 structure). Dashed line is to highlight the nine spins 
 unit cell. 
}
\label{cartoon}
\end{figure}

In this Article we consider the two dimensional spin-$\frac{1}{2}$ 
Heisenberg model with both nearest and next-nearest 
neighbor interactions ($J_1$-$J_2$), defined by the $SU(2)$ symmetric 
Hamiltonian 
\begin{equation}
\label{j1j2_ham}
{\mathcal H}=J_1\sum\limits_{\langle ij\rangle}\mathbf{S}_i
\cdot \mathbf{S}_j+J_2\sum\limits_{\langle\langle i,k
\rangle\rangle}\mathbf{S}_i\cdot\mathbf{S}_k.
\end{equation}
Here $\mathbf{S}_i$ are spin-$\frac{1}{2}$ operators and 
$\langle i,j\rangle$, $\langle\langle i,k\rangle\rangle$ denote   
respectively nearest neighbor and next-nearest neighbor sites on 
the lattice.  We consider both triangular and kagom\'e cylinders 
of size $W\times L$ (Figure~\ref{cartoon} (i) and (iii) respectively) 
with periodic boundary conditions along the vertical direction. We choose  
$J_1>0$ (antiferromagnetic nearest neighbors interactions) and $J_2<0$ 
({\it ferromagnetic} next-nearest neighbors interaction). Clearly, a large 
negative $J_2$ favors the formation of ferromagnetic sublattices (cf. 
Figure~\ref{cartoon} (i)(iii)) and magnetic order~\cite{lhuillier_arxiv2005}. 
Here, in particular, we restrict ourselves to $J_2/J_1=-1$ to ensure 
a magnetically ordered ground state on both the triangular and 
kagom\'e lattice.

\paragraph{The triangular lattice.---}

The ground state of the $J_1$-$J_2$ Heisenberg model on the triangular 
lattice (THA) exhibits at $J_2/J_1=-1$ (at a semiclassical 
level, i.e. considering large spins $S\gg 1/2$) the so called 
$120^\circ$ structure. This is depicted in Figure~\ref{cartoon} (ii)  
and consists of three ferromagnetic sublattices (associated with full 
breaking of spin rotational invariance). Spins on the same 
sublattice are parallel, while the angle between spins in different 
sublattices is $120^\circ$. A possible choice of ordering pattern  
is shown in Figure~\ref{cartoon} (spin orientations are denoted 
as $\alpha,\beta,\gamma$). For spins $S=1/2$ (which is the case of 
interest here) quantum fluctuations are not strong enough to destroy the 
magnetic order and the $120^\circ$ structure survives. One should mention 
that this remains true at arbitrary $J_2\le0$, as confirmed by spin-wave 
calculations~\cite{oguchi-1983,jolicoeur-1989,miyake-1992,singh-1992,chubukov-1994,
chernyshev-2009}, Green's function Monte Carlo~\cite{capriotti-1999}, series 
expansions~\cite{zheng-2006}, tower of states spectroscopy~\cite{bernu-1992}, 
and recent DMRG calculations~\cite{white-2007}.

\paragraph{The kagom\'e lattice.---}

Much less is known about the phase diagram of the $J_1$-$J_2$ Heisenberg 
model on the kagom\'e lattice (KHA) (cf. Figure~\ref{cartoon} (iii)). 
At large ferromagnetic $J_2<0$ (in particular at $J_2/J_1=-1$) the ground state 
exhibits  magnetic order {\it \`a la N\'eel} with spontaneous breaking of 
$SU(2)$ symmetry. The selected ordering pattern  is the  
$\sqrt{3}\times\sqrt{3}$ state (cf. Figure~\ref{cartoon} (iv)). As for the THA 
(Figure~\ref{cartoon} (ii)), three ferromagnetic sublattices are present,  
although the unit cell (highlighted with the dashed line in the Figure) is 
now larger (it contains nine spins).

One should mention that, while it is well established that the $\sqrt{3}\times\sqrt{3}$ 
order survives at smaller $J_2$ (i.e. at $J_2>-1$)~\cite{lech-1997},  it is still a 
challenging task to determine the phase diagram of the $J_1$-$J_2$ KHA in the limit 
$J_2\approx 0$. In particular, the nature of the ground state of the pure kagom\'e 
Heisenberg antiferromagnet (i.e. at $J_2=0, J_1>0$)  is still debated. 
Several valence bond crystals~\cite{marston-1991,zeng-1990,hastings-2000, 
nikolic-2003,singh-2007,singh-2008,iqbal-2012} and spin liquid ground 
states~\cite{messio-2012,yan-1993, hermele-2008,iqbal-2011,ran-2007,ryu-2007, 
sachdev-1992,wang-2006,lu-2011,misguich-2002,jiang-2008,huh-2011,kalmeyer-1989} 
(both gapless and gapped) have been proposed. Remarkably, recent state-of-the-art DMRG 
calculations have provided robust evidence of a gapped $Z_2$ topological 
spin liquid~\cite{white-2011,depenbrock-2012}. Interestingly, there is also 
evidence that the spin liquid behavior might survive at small positive $J_2$ 
with the formation of an extended spin liquid region~\cite{jiang-2012}.

\paragraph{Entanglement spectrum (ES).---} 

In order to calculate the ES we consider the bipartition of the system 
(cylinders in Figure~\ref{cartoon}) into two equal parts $A$ and $B$, 
using a vertical cut (dashed line along the $y$-direction in  
Figure~\ref{cartoon} (i)(iii)). As a consequence, the boundary between 
$A$ and $B$ is a circumference of length $W$. The  
total subsystem  spin $\mathbf{S}^2_A$ is a good ``quantum number'' 
for the ES and  can be used to label ES levels (i.e. ES levels 
are organized into $SU(2)$ multiplets). Equivalently, the entanglement 
Hamiltonian ${\mathcal H}_E$ (or the reduced density matrix $\rho_A$) 
exhibits a block structure, each block corresponding 
to a different $\mathbf{S}_A$ sector.  

\paragraph{Ground state search (DMRG method).---}

The ground state is obtained in a matrix product state form by using 
state-of-the-art $SU(2)$-symmetric single-site DMRG \cite{White_1992, 
Schollwock_2005, Schollwock_2011}. DMRG (Density Matrix Renormalization Group) 
is a variational method in the ansatz space spanned by matrix product states (MPS). 
The method allows one to find the ground state of one-dimensional (1D) systems 
efficiently even for large system sizes. It has also been successfully applied to 
two-dimensional (2D) lattices by mapping the short-ranged 2D Hamiltonian 
exactly to a long-ranged 1D one~\cite{White_spin-gaps_1996, 
White_tj_1998, White_energetics_1998, white-2007, white-2011, 
stoudenmire_studying_2011, depenbrock-2012}. Here, to ensure independence 
on the actual mapping, we performed several calculations using different 
mappings. DMRG computational cost scales roughly exponentially with 
the entanglement entropy and favors open (OBC) over periodic boundary conditions 
(PBC). The conventional compromise, taken also by us, is to consider cylinders, i.e.  
PBC along the short direction (circumference $W$) and OBC along the long direction 
(length $L$) where boundary effects are less important. Computational cost  
is then dominated exponentially by $W$. Exploiting the power of the non-abelian 
formulation we were able to simulate the systems using up to 5,000 ansatz 
states, corresponding to roughly 20,000 states in an abelian $U(1)$ DMRG, 
allowing us to obtain the ground state of~\eqref{j1j2_ham} with high accuracy, 
even for cylinders with $W=9$ (for the $J_1$-$J_2$ THA) or fully periodic tori. 
One should also mention that in $SU(2)$-broken phases the large entanglement gap, 
which  divides  the TOS ES levels from the rest, reduces significantly the 
effective number of states needed to get well converged ground states.

\section{Tower of states spectroscopy in SU(2) broken phases}
\label{energy_tos}

\begin{figure*}[t]
\includegraphics[width=.9\linewidth]{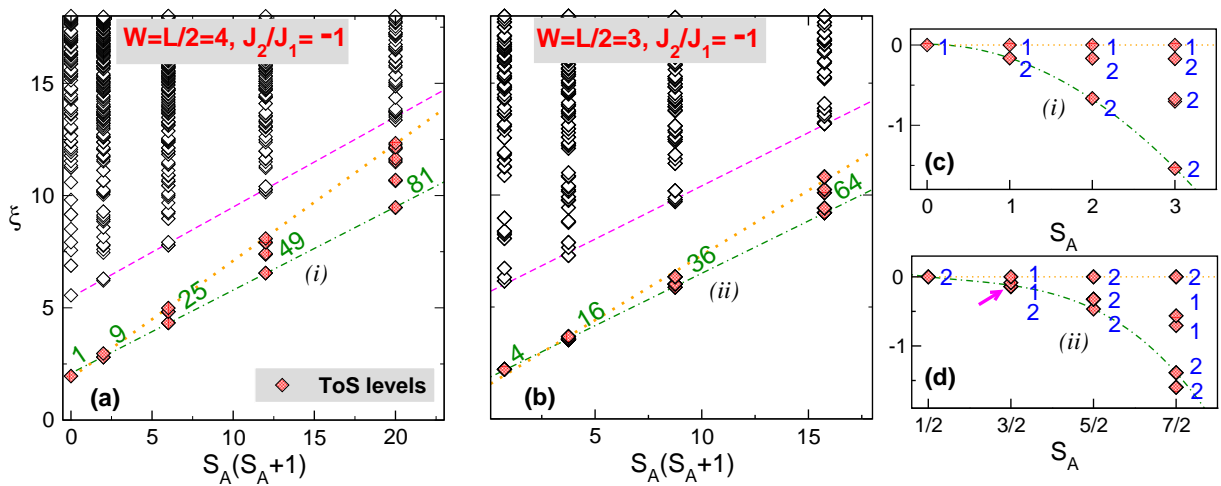}
\caption{ Tower of states (TOS) structure in the ES 
 of the $J_1$-$J_2$ kagom\'e Heisenberg model (KHA) at $J_2/J_1=-1$. Half-system 
 ES levels $\xi$ versus $S_A(S_A+1)$, with $S_A$ the block spin. 
 Symbols are DMRG data for the KHA on a cylinder with $W=L/2=4$ (cf. (a)) 
 and $W=L/2=3$ (cf. (b)), same scale  used on the $y$-axis. Each point 
 corresponds to a degenerate $SU(2)$ multiplet ($2S_A+1$ levels). Filled 
 symbols denote levels building the TOS. Dashed-dotted line is to highlight 
 the behavior as $S_A(S_A+1)$. TOS levels are divided from the rest 
 (levels above the dashed line in the Figure) by an entanglement 
 gap. Accompanying numbers are the numbers of TOS ES levels. 
 Right panels (c)(d): Enlarged view of the TOS structures {\it (i)(ii)} 
 shown respectively in (a) and (b). In each $S_A$ sector ES levels 
 are shifted by the value of the highest level (dotted line in (a)(b)). 
 ES levels are plotted against the block spin $S_A$. The number of 
 degenerate $SU(2)$ multiplets is reported in blue. In (d) the arrow is to 
 stress the presence of isolated (i.e. unpaired) multiplets (see also 
 multiplets at $S_A=7/2$).
}
\label{fig1_tos_ex}
\end{figure*}

Due to its manifest spin rotational invariance, the finite size 
spectrum of~\eqref{j1j2_ham} can be decomposed into the irreducible 
representations of $SU(2)$, using the eigenvalue $S$ of the total spin 
$\mathbf{S}^2$ to label energy levels (and eigenstates). 
The resulting spin-resolved spectrum shows striking signatures of the $SU(2)$ 
breaking (happening in the thermodynamic limit). Exact 
diagonalization studies~\cite{lech-1997} demonstrated that at 
$J_2/J_1=-1$ in each spin sector $S$ there is a family of (low-lying) 
levels, which are clearly separated from the rest by an energy gap 
(at least for large systems). These are called ``quasidegenerate joint 
states'' (QDJS) in Ref.~\cite{bernu-1994} and form the ``tower of states'' 
(TOS) structure. 

The number $N_S$ of TOS levels in each spin sector is related to the  
N\'eel state selected in the thermodynamic limit. For instance,  
N\'eel ordering with two ferromagnetic sublattices (as for the Heisenberg 
antiferromagnet on the square lattice~\cite{lhuillier_arxiv2005}), 
corresponding to the breaking of $SU(2)$ down to $U(1)$, implies 
$N_S=2S+1$. On the other hand, a complete breaking of $SU(2)$ (for instance  
N\'eel ordering with more than two ferromagnetic sublattices, as for both 
the THA and KHA, cf. Figure~\ref{cartoon}) implies 
$N_S>2S+1$~\cite{lhuillier_arxiv2005} (see also below).

The TOS structure can be obtained as the lowest energy manifold of an 
effective Hamiltonian ${\mathcal H}_{T}$ (``quantum top''), which, for 
N\'eel order with three ferromagnetic sublattices $a,b,c$, 
reads~\cite{anderson,bernu-1992,siggia-1989,fisher-1989,azaria-1993}
\begin{eqnarray}
\label{TOS_ham}
{\mathcal H}_{T}=\frac{1}{2\chi V}(\mathbf{S}^2-\mathbf{S}_a^2-
\mathbf{S}^2_b-\mathbf{S}_c^2)\equiv \\\nonumber\frac{1}{\chi V}
(\mathbf{S}_a\cdot\mathbf{S}_b+\mathbf{S}_a\cdot\mathbf{S}_c+\mathbf{S}_b 
\cdot\mathbf{S}_c).
\end{eqnarray}
Here $\chi$ is the spin susceptibility, $V$ the volume (i.e. total number of sites), 
and $\mathbf{S}$ ($\mathbf{S}_{a,b,c}$)  the total spin of the system (sublattice). 
Notice that one could think of~\eqref{TOS_ham} as an effective Heisenberg 
coupling between $\mathbf{S}_a,\mathbf{S}_b,\mathbf{S}_c$, acting as collective 
degrees of freedom. As the lowest energy manifold of~\eqref{TOS_ham} is obtained 
choosing $S_a=S_b=S_c=V/3\times 1/2$, one readily obtains 
the number of TOS levels per spin sector as $N_S=(2S+1)^2$~\cite{bernu-1994}. 
These, according to~\eqref{TOS_ham}, are degenerate with energy given as 
\begin{equation}
\label{TOS_spectrum}
E_{T}(S)=\frac{1}{2\chi V}S(S+1),
\end{equation}
where we neglected the sublattice contributions, keeping only $S$ dependent terms. 
Plotted as function of $S(S+1)$, TOS levels show the typical ``Pisa tower'' (linear) 
structure~\cite{bernu-1994},  with a  vanishing (as $1/V$) 
``slope''.

Still, one should think of~\eqref{TOS_ham} only as the low energy 
approximation of~\eqref{j1j2_ham}. To go beyond one can  
split  ${\mathcal H}$ as ${\mathcal H}={\mathcal H}_{T}+{\mathcal H}'$
with ${\mathcal H}'$ a (higher energy) correction to ${\mathcal H}_{T}$. 
Specifically, one has ${\mathcal H}'\approx {\mathcal H}_{sw}$,   
${\mathcal H}_{sw}$ describing levels immediately above the TOS 
structure. These correspond to spin waves (Goldstone modes) and possess 
a linear dispersion, implying (using that momentum is discretized on a 
finite lattice as $1/\sqrt{V}$) ${\mathcal H}_{sw}\approx 1/\sqrt{V}$. 
As a striking consequence the TOS spectrum~\eqref{TOS_spectrum} is divided 
from higher energy levels by an apparent gap at $V\to\infty$. 

Moreover, since in general $[{\mathcal H}_{T},{\mathcal H}']\ne 0$,   
the degeneracy within each TOS manifold at spin $S$ (cf.~\eqref{TOS_spectrum}) 
is partly lifted, implying that ${\mathcal H}_T$ (cf.~\eqref{TOS_ham}) has to be 
modified. Notice that, in principle, the final degeneracy structure can be 
predicted using group symmetry analysis~\cite{bernu-1994}. Remarkably, in 
the limit of large systems  ${\mathcal H}_{T}$ can be mapped 
to the anisotropic ``quantum top''~\cite{bernu-1994}
\begin{equation}
\label{tos_imp}
{\mathcal H}_{T}=\frac{\mathbf{S}^2}{2V\chi_\perp}+\frac{(S^{z'})^2}
{2V}\left(\frac{1}{\chi_\parallel}-\frac{1}{\chi_\perp}\right).
\end{equation}
Here $S^{z'}\in [-S,S]$ is the component of the total spin along the 
third axis $z'$ of the ``quantum top'' (not necessarily the $z$ axis 
in the lab-frame), while $\chi_\parallel$ and $\chi_\perp$ denote 
respectively the parallel  and transverse susceptibilities, which 
measure the response to magnetic fields in the plane of the spins 
and in the perpendicular one. Notice that  both terms in~\eqref{tos_imp} 
are $\sim 1/V$. One has for large system sizes 
$\chi_\perp\ne\chi_\parallel$, reflecting the tendency towards 
magnetic order and  the system response becoming  anisotropic. 
The degeneracy structure of TOS multiplets is now readily obtained 
from~\eqref{tos_imp}: in the sector with half-integer 
$S$ there are $S+1/2$ pairs  of degenerate multiplets, whereas for 
integer $S$ one has $S$ degenerate pairs and an extra isolated multiplet 
(corresponding to $S^{z'}=0$ in~\eqref{tos_imp}).

\section{Entanglement spectra in SU(2)-broken phases}
\label{tos_es}

\begin{figure*}[t]
\includegraphics[width=.9\linewidth]{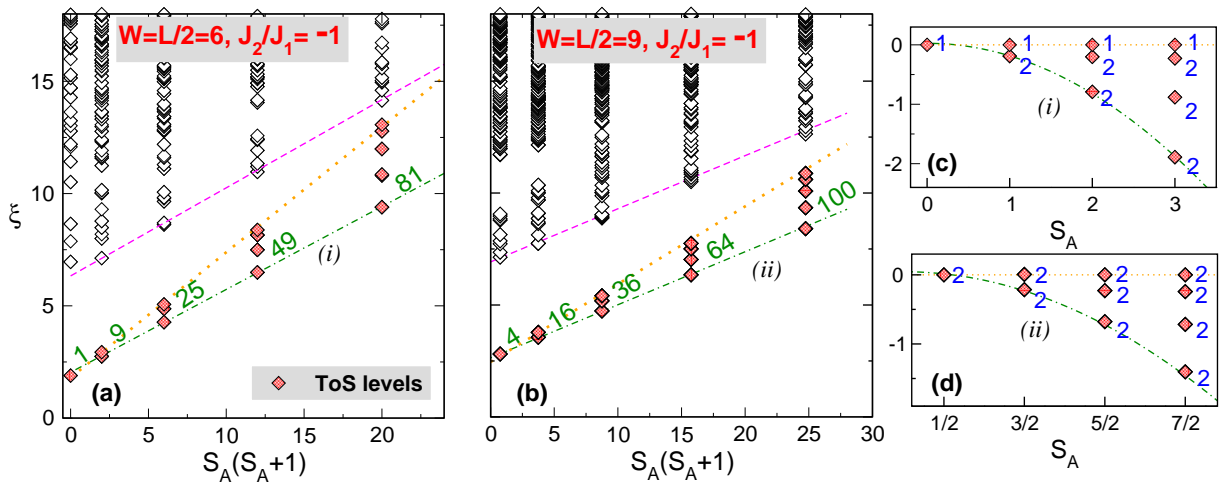}
\caption{ Tower of states (TOS) structure in the ES of the $J_1$-$J_2$ 
 Heisenberg model on the triangular lattice (THA) ($J_2/J_1=-1$). 
 ES for half of the system:  ES levels $\xi$ versus $S_A(S_A+1)$,  
 $S_A$ being the total subsystem spin. Symbols are DMRG data for cylinders 
 with $W=L/2=6$ (a) and $W=L/2=9$ (b) (cf. Figure~\ref{cartoon}). 
 Each point corresponds to a degenerate $SU(2)$ multiplet ($2S_A+1$ levels). 
 Filled symbols denote the ES levels forming the TOS.  Dashed-dotted line 
 highlights the TOS behavior as $S_A(S_A+1)$. TOS levels are divided from 
 the rest of the spectrum (levels above the dashed line) by an entanglement gap. 
 The total number of ES levels in each $S_A$ sector is reported in green 
 (numbers accompanying ES multiplets). 
 Right panel: Enlarged view of the TOS structures in (a) and (b), ES plotted 
 versus $S_A$. ES levels at each $S_A$ are shifted by the highest level 
 (dotted lines in (a)(b)). Lines are guides to the eye as in (a)(b). 
 Accompanying numbers denote the number of degenerate multiplets.
}
\label{fig1_tos_ex_T}
\end{figure*}

In this section we numerically demonstrate that in $SU(2)$-broken 
phases the information contained in the energy tower of states (TOS) 
is nicely embodied in the lower part of the ground state entanglement spectrum 
(TOS-ES correspondence). This section is organized as follows. In~\ref{TOS-ES} 
we establish the TOS-ES correspondence~\cite{met-grov-2011,alba-2012}, which 
is expressed as a mapping between the TOS Hamiltonian ${\mathcal H}_T$ and 
the entanglement Hamiltonian ${\mathcal H}_E$. This is supported numerically 
in \ref{kagome-ES} highlighting TOS structures in the ES of the $J_1$-$J_2$ 
KHA and THA (at $J_2/J_1=-1$). Our main results are illustrated in 
Figure~\ref{fig1_tos_ex} and~\ref{fig1_tos_ex_T}. Finally, the fine  
structure (TOS substructure) of the TOS-related levels is detailed  
in~\ref{asy_top}. 

\subsection{TOS-ES correspondence}
\label{TOS-ES}

It has been suggested recently that in systems breaking a {\it continuous}  
symmetry in the thermodynamic limit the lower part of the (ground state) 
ES has the same structure as the TOS  energy spectrum~\cite{met-grov-2011}. 
Here we restrict ourselves to the situation of $SU(2)$ symmetry breaking. 
The correspondence can be expressed as a mapping between an effective 
entanglement Hamiltonian ${\mathcal H}_E$ (describing the lower part 
of the ES) and the TOS Hamiltonian ${\mathcal H}_{T}$. Specifically, 
one has~\cite{met-grov-2011} 
\begin{equation}
\label{tos_ent}
{\mathcal H}_E\propto {\mathcal H}_{T}(A)/T_E,
\end{equation}
where ${\mathcal H}_{T}$ is restricted to the degrees of freedom of 
subsystem A and $T_E\approx v_s/\sqrt{V}$ is an effective ``entanglement 
temperature'', which reflects the presence of gapless excitations (spin waves) 
arising from the breaking of the $SU(2)$ symmetry (here $v_s$ is the spin 
wave velocity). The behavior $T_E\approx 1/\sqrt{V}$ originates from the 
linear dispersion of spin waves and  the momentum quantization as $1/\sqrt{V}$ 
on a finite lattice.

From~\eqref{tos_ent} two remarkable properties can be derived. First, using 
that ${\mathcal H}_{T}\sim 1/V$ (cf.~\eqref{tos_imp}) and $T_E\sim 1/\sqrt{V}$, 
one obtains that the spacing between the ES levels building the TOS structure 
is vanishing as $1/\sqrt{V}$ in the thermodynamic limit. 
Additionally, including the spin wave contributions in the energy spectrum, 
i.e. replacing ${\mathcal H}_{T}\to {\mathcal H}_{T}+{\mathcal H}_{sw}$, 
and assuming that ES levels above the TOS structure are spin wave like, 
from~\eqref{tos_ent} one obtains ${\mathcal H}_E$ as  
\begin{equation}
{\mathcal H}_E\sim ({\mathcal H}_{T}+{\mathcal H}_{sw})/T_E.
\end{equation}
The behaviors $T_E\sim 1/\sqrt{V}$ and ${\mathcal H}_{sw}\sim 1/\sqrt{V}$ 
now suggest the formation of a finite gap (in the limit $V\to\infty$)  
between the TOS structure and the higher part 
of the ES. However, one should stress that a logarithmic vanishing of the 
entanglement gap, also suggested by field theoretical 
calculations~\cite{private}, cannot be excluded. These findings (presence 
of a finite gap in the ES and the finite size behavior of the TOS structure) 
have been confirmed in Ref.~\cite{alba-2012} for the 2D Bose-Hubbard model 
in the superfluid phase. 

Finally, it is interesting to discuss how TOS structures affect the 
behavior of the entanglement entropy. The fact that 
the low-energy part of~\eqref{j1j2_ham} (and its ground state) can be 
described by an effective free bosonic theory (${\mathcal H}_{sw}$, 
cf. section~\ref{energy_tos}) suggests that an area law behavior should be  
expected (cf.~\cite{amico-2008} and references therein for a discussion of 
area laws in free systems). On the other hand, the breaking of a continuous 
symmetry gives rise to additive logarithmic corrections to the 
entropy~\cite{met-grov-2011}, which, for instance, have been observed numerically 
in the 2D Heisenberg antiferromagnet on the square lattice~\cite{kallin-2009,
hastings-2010,song-2011}. At the level of the ES, these corrections are 
associated with the TOS structure, while the area law arises from ES levels 
above the entanglement gap. Note that the entanglement gap is typically large 
deep in a $SU(2)$-broken phase (see section~\ref{kagome-ES}), implying that  the 
TOS levels give the dominant contribution to the entanglement entropy, while 
the area law behavior is recovered only asymptotically for large system sizes.

\subsection{DMRG results}
\paragraph*{$J_1$-$J_2$ kagom\'e Heisenberg (KHA).---}
\label{kagome-ES}

We start discussing the tower of states structures in the ES of  
the KHA at $J_2/J_1=-1$. Figure~\ref{fig1_tos_ex} plots the ES (DMRG data)
obtained from the ground state of the KHA on  cylinders (cf. Figure~\ref{cartoon} 
(iii)) with fixed aspect ratio $W/L=1/2$ and $W=3,4$ (respectively center and left  
panels in Figure~\ref{fig1_tos_ex}). Total number of spins in the subsystem 
is given as $3W^2$ (ES is for half cylinder) and is even (odd) for $W=4(3)$. 
ES levels $\xi$ are plotted versus $S_A(S_A+1)$, $S_A$ being the total spin 
of subsystem $A$. 

In each spin sector $S_A$ a family of low-lying ES multiplets 
(each point corresponds to an $SU(2)$ multiplet of degenerate levels, 
filled rhombi in Figure~\ref{fig1_tos_ex}) is well separated from higher levels 
by a gap. The total number  of levels below the gap  (TOS levels) 
in each sector $S_A$ is given as $(2S_A+1)^2$ (numbers 
accompanying ES multiplets in the Figure), clearly reflecting the 
corresponding multiplicity (as $(2S+1)^2$) in the energy tower of 
states (cf. section~\ref{energy_tos}). Also, the lower part of the 
TOS  levels exhibits the typical TOS behavior as $S_A(S_A+1)$ (see 
dashed-dotted lines in Figure~\ref{fig1_tos_ex}) in agreement 
with~\eqref{TOS_spectrum} and~\eqref{tos_ent}. The entanglement gap 
appears to be constant as a function of $S_A$ (dashed line denotes levels 
immediately above the TOS structure), similarly to what  is observed in
energy TOS structures~\cite{lhuillier_arxiv2005} and in the ES of 
the 2D Bose-Hubbard~\cite{alba-2012}.

Interestingly, at each fixed $S_A$ the TOS levels are not exactly degenerate,  
and further substructures appear, reflecting the presence of the second 
term in~\eqref{tos_imp}. TOS substructures are better highlighted in 
Figure~\ref{fig1_tos_ex} (c)(d) showing an enlarged view of 
the TOS levels (same DMRG data as in panels (a)(b)). In each sector 
$S_A$ we shifted the ES by subtracting the value of the largest 
level (dashed-dotted and dashed lines are guides to the eye as in panels 
(a)(b)). Reported numbers correspond now to the number of degenerate $SU(2)$ 
multiplets. 

According to~\eqref{tos_ent} the degeneracy structure in the TOS part of the ES
is the same as that in the energy tower of states. At large system sizes and 
{\it integer} $S_A$ (i.e. even number of spins in $A$) the TOS ES  levels  
are organized in pairs of degenerate multiplets, apart from one isolated 
multiplet at the top of each $S_A$ sector. This is clearly supported in 
Figure~\ref{fig1_tos_ex} panel (c).

\begin{figure}[t]
\includegraphics[width=.95\columnwidth]{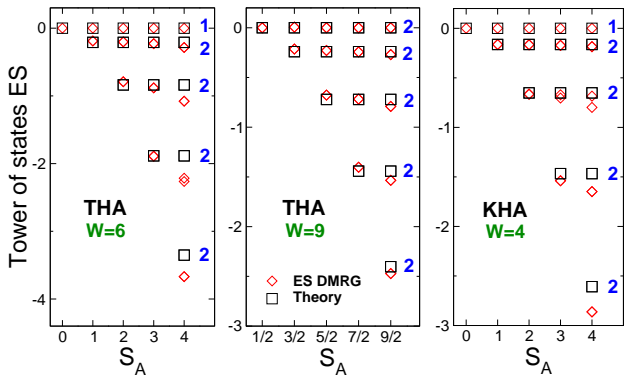}
\caption{ TOS entanglement substructures. ES of the Heisenberg 
 $J_1$-$J_2$ model on the triangular (THA) and kagom\'e (KHA) 
 lattice: ES levels $\xi$ versus the total spin $S_A$ of 
 subsystem $A$. Each point corresponds to a degenerate 
 $SU(2)$ multiplet ($2S_A+1$ levels). Only multiplets building 
 the TOS structure are shown. In each spin sector with fixed 
 $S_A$ ES multiplets are shifted by the value of the largest level. 
 Rhombi are the same DMRG data as in  Figure~\ref{fig1_tos_ex} 
 and~\ref{fig1_tos_ex_T} panels (c)(d)). The squares denote 
 the (one parameter) fit to the theoretical prediction 
 (cf.~\eqref{tos_asy}) in the limit of large systems 
 ($W,L\to\infty$). In all panels accompanying numbers 
 denote the number of degenerate $SU(2)$ multiplets.
 }
\label{fig6_top}
\end{figure}

On the other hand, for {\it half-integer} 
$S_A$ only pairs of degenerate multiplets are expected (cf.~\eqref{tos_imp}). 
Figure~\ref{fig1_tos_ex} (d) shows the TOS ES levels for the kagom\'e cylinder  
with $W=3$ (i.e. 27 spins in subsystem $A$). Although a clear tendency towards 
the formation of pairs is visible (levels at the top of the structure form  pairs, 
while in panel (c) one has one isolated multiplet), some deviations are observed. 
For instance (see arrow in Figure~\ref{fig1_tos_ex} (d)), one has in the 
sector with $S_A=3/2$  four $SU(2)$  multiplets, but only two form a pair. 
Similarly,  in the sector with $S_A=7/2$ two isolated multiplets are 
visible. Since~\eqref{tos_imp} is valid only in the asymptotic (i.e. large $V$) 
regime, these deviations have to be understood as finite size effects. Indeed,
we checked that at $W=5$ (i.e.  $75$ spins in subsystem $A$) all the (TOS)   
multiplets form degenerate pairs (at least in the first few $S_A$ sectors).

\paragraph*{$J_1$-$J_2$ triangular Heisenberg (THA).---}
\label{triangular-ES}

Further evidence supporting the TOS-ES scenario is provided in 
Figure~\ref{fig1_tos_ex_T} considering the 2D $J_1$-$J_2$ Heisenberg model 
on the triangular lattice (THA). The ground state ordering pattern 
($120^\circ$ structure, cf. Figure~\ref{cartoon} (ii)) contains three 
ferromagnetic sublattices (full breaking of $SU(2)$) and the same tower of 
states structure as for the kagom\'e  is expected.

Figure~\ref{fig1_tos_ex_T} plots DMRG data for the ES of the THA on 
the cylinder (at fixed aspect ratio $W/L=1/2$ with $W=6$ and $W=9$, 
respectively in panel (a) and (b)). ES is for half of the cylinder. Notice 
that we could access larger system sizes than for the kagom\'e (compare with 
Figure~\ref{fig1_tos_ex}). This allows us to resolve the TOS multiplets 
(corresponding to $100$ ES levels) at $S_A=9/2$. As for the kagom\'e ES 
(cf. Figure~\ref{fig1_tos_ex})  the lower part of the ES (filled symbols 
in the Figure) is divided from the rest of the spectrum by an entanglement 
gap and exhibits the typical TOS behavior as $S_A(S_A+1)$.

The correct $SU(2)$ TOS level counting (i.e. number of TOS levels in each 
spin sector $S_A$) as $(2S_A+1)^2$ is fully reproduced. The fine structure  
of TOS multiplets (TOS substructure) is highlighted in Figure~\ref{fig1_tos_ex_T} 
(c)(d). Remarkably, for odd number of spins  in $A$ all TOS levels are 
organized into pairs of degenerate multiplets (cf. Figure~\ref{fig1_tos_ex_T} 
(d)), whereas for even ones there is an isolated ES multiplet at the top of 
the structure (cf. Figure~\ref{fig1_tos_ex_T} (c)), signaling that finite 
size corrections are somehow smaller than in the kagom\'e ES (cf. 
Figure~\ref{fig1_tos_ex}).

\subsection{Tower of states entanglement substructures}
\label{asy_top}

We now analyze quantitatively the structure of the TOS ES multiplets. 
We start with observing that in the limit of large cylinders the effective 
entanglement Hamiltonian ${\mathcal H}_E$ describing the TOS structure  
is obtained from~\eqref{tos_imp} and~\eqref{tos_ent}  as 
\begin{equation}
\label{tos_asy}
{\mathcal H}_{E}\sim \frac{\mathbf{S}_A^2}{v_s\chi_\perp W}-
\frac{(S^{z'}_A)^2}{v_s W}\Big(\frac{1}{\chi_\perp}-\frac{1}
{\chi_\parallel}\Big),
\end{equation}
where we used that $\sqrt{V}\approx W$. While the first term 
in~\eqref{tos_asy} gives the TOS behavior as $S_A(S_A+1)$ 
(cf. Figure~\ref{fig1_tos_ex} and~\ref{fig1_tos_ex_T}), with  
$(2S_A+1)^2$ {\it degenerate} levels at each $S_A$, the second 
gives rise to the substructures in Figure~\ref{fig1_tos_ex} 
and~\ref{fig1_tos_ex_T} (c)(d).

These are shown in  Figure~\ref{fig6_top} plotting the shifted ES levels 
(same DMRG data as in Figure~\ref{fig1_tos_ex} and~\ref{fig1_tos_ex_T} 
panels (c)(d)) for both the triangular and kagom\'e $J_1$-$J_2$ Heisenberg 
model at $J_2/J_1=-1$.  Since ES levels in each sector $S_A$ are shifted 
by the value of the largest level, the contribution of the first term 
($\sim S_A(S_A+1)$) in~\eqref{tos_asy} has to  be neglected. Thus, structures 
appearing in Figure~\ref{fig6_top} are described by $\alpha [(S_A^{z'})^2+s_0]$, 
being $\alpha\sim (\chi_\parallel-\chi_\perp)/(v_sW\chi_\perp\chi_\parallel)$, 
and $s_0=0(-1/4)$ for integer(half-integer) values of $S_A$. 

This scenario is confirmed fitting TOS levels in Figure~\ref{fig6_top} to 
$\alpha [(S_A^{z'})^2+s_0]$, with $\alpha$ the only fitting parameter. 
For the THA  (including in the fit only the ES towers with $S_A\le 3$) 
it is  $\alpha\approx 0.21$, while for $W=9$ (now including all the 
ES levels with $S_A\le 9/2$) one obtains $\alpha\approx 0.12$. 
Notice that it is $0.12/0.21\sim 0.6\sim 2/3$, supporting the behavior 
$\alpha\sim 1/W$ (cf.~\eqref{tos_asy} and section~\ref{fin_size}). For 
the KHA ($W=4$) a similar fit gives $\alpha\approx 0.17$, (only ES levels 
with $S_A\le 3$ were fitted). Results of the fit are shown in Figure~\ref{fig6_top} 
as squares and are in excellent agreement with the DMRG data. Also, the 
agreement is better at larger system sizes (compare in Figure~\ref{fig6_top} DMRG 
data  for the THA at $W=9$ and $W=6$), confirming that~\eqref{tos_asy} holds 
in the asymptotic regime $V\to\infty$.

\section{Finite size and boundary effects in TOS structures}
\label{fin_size}

One crucial consequence of the correspondence between TOS and entanglement 
spectra, according to~\eqref{tos_asy}, is that the spacing within low-lying 
ES multiplets  is $\sim1/\sqrt{V}\approx 1/W$. Oppositely, 
the entanglement gap between the TOS part and the rest of the spectrum 
remains finite in the thermodynamic limit (or vanishes logarithmically, 
cf. the discussion in section~\ref{TOS-ES}). These features are numerically 
demonstrated in~\ref{gaps}.

The effect of boundary conditions on TOS structures is instead discussed 
in~\ref{boundaries}, by examining the ES of the $J_1$-$J_2$ KHA on the torus. 
The most notable consequence of the torus geometry is that the number of 
boundaries between the two subsystems is doubled. However, although this 
gives rise to quantitative differences compared  to the cylinder geometry, 
qualitative features (i.e. TOS behavior as $S_A(S_A+1)$ and TOS multiplets 
counting) remain unchanged, signaling the bulk origin of the TOS structures.

\subsection{Entanglement gap \& TOS level spacing: finite size scaling 
analysis}
\label{gaps}

\begin{figure}[t]
\includegraphics[width=1\columnwidth]{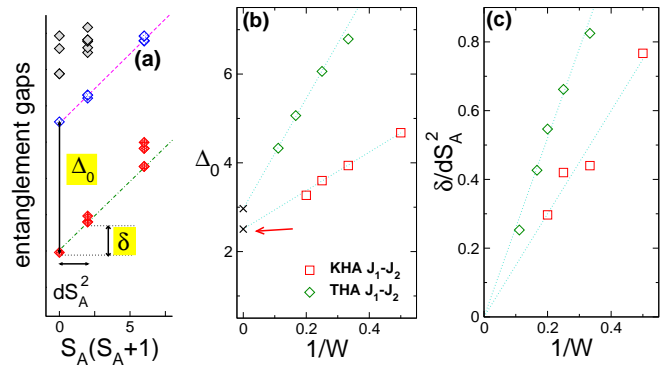}
\caption{ Finite size scaling of the entanglement gap $\Delta_0$  
 and the tower of states spacing $\delta$ in the $J_1$-$J_2$ Heisenberg 
 model on the kagome (KHA) and triangular (THA) lattice (at $J_2/J_1=-1$). 
 DMRG data for the ES of half the system (cylindrical geometry 
 as in  Figure~\ref{cartoon} with fixed aspect ratio $W/L=1/2$). (a) 
 Pictorial definitions of entanglement gap and tower level spacing: 
 $\delta$ is the ``distance'' between the two lowest levels in 
 the ES (here respectively in the sectors $S_A=0$ and $S_A=1$). 
 $\Delta_0$ is the gap between the TOS structure and higher ES levels 
 in the $S_A=0$ sector. (b) $\Delta_0$ as 
 function of $1/W$: the gap is finite in the limit $W\to\infty$. Dotted 
 lines are fits to $A+B/W$. The extrapolated values for $A$ are shown 
 as crosses. (c) Vanishing of the tower level spacing $\delta$ in the 
 thermodynamic limit. To avoid odd-even effects $\delta$ is divided by
 $dS_A^2=2,3$ for respectively integer and half-integer $S_A$. 
 $\delta/dS^2_A$ plotted versus  $1/W$. Dotted lines are fits to 
 $A/W$.
}
\label{fig3_TOS_struct}
\end{figure}

The structure of the lower part of the ES (TOS structure) 
can  be characterized using the entanglement gap $\Delta_0$ 
and the tower of states level spacing $\delta$~\cite{alba-2012}.
These are defined pictorially in Figure~\ref{fig3_TOS_struct} (a).
More formally, $\delta$ is the ``distance'' between the two lowest 
levels in the sectors with  $S_A=0,1$ (respectively $S_A=1/2,3/2$ 
for $S_A$ half integer), i.e.   $\delta\equiv\xi_{1}-\xi_{0}$ 
with $\xi_{\sigma}$ the lowest ES level in the sector with 
$S_A=\sigma$. This is also a measure of the ``slope'' of 
the TOS structure. The entanglement gap  $\Delta_0$ measures, instead, 
the separation between the TOS structure and the higher ES levels. 
Since  it depends weakly on $S_A$ (cf. Figure~\ref{fig1_tos_ex} 
and~\ref{fig1_tos_ex_T}), here we consider  the gap $\Delta_0$ 
in the lowest spin sector ($S_A=0(1/2)$ for integer (half-integer) 
$S_A$).  

Figure~\ref{fig3_TOS_struct} (b) plots $\Delta_0$ as function 
of the boundary length $2\le W\le 9$ for both the kagom\'e and 
triangular $J_1$-$J_2$ Heisenberg model ($J_2/J_1=-1$). The ES is 
for half of the system and data is DMRG for cylinders with fixed 
aspect ratio $W/L=1/2$. For both models the extrapolation to infinite 
cylinders (assuming the behavior $1/\sqrt{V}\sim 1/W$) (dotted lines) 
suggests a finite value (crosses in the Figure) of $\Delta_0$ 
(see, however, the discussion in~\ref{TOS-ES}). 

Figure~\ref{fig3_TOS_struct} (c) shows $\delta$ versus $1/W$. In order
to avoid parity effects (in $V$) we plot $\delta/dS^2_A$, with 
$dS^2_A\equiv \left. S_A(S_A+1)\right |_{1}-\left. S_A(S_A+1)\right 
|_{0}$. Clearly,  this is vanishing for infinite cylinders 
($W\to\infty$). The expected behavior $\delta\sim 1/\sqrt{V}\sim 1/W$ 
(cf.~\eqref{tos_asy}) is fully confirmed for the $J_1$-$J_2$ THA  (rhombi in 
the Figure, dotted line is a fit to $A/W$), while for the $J_1$-$J_2$ KHA 
the scenario is less robust due to residual parity effects.

\begin{figure}[t]
\includegraphics[width=.85\columnwidth]{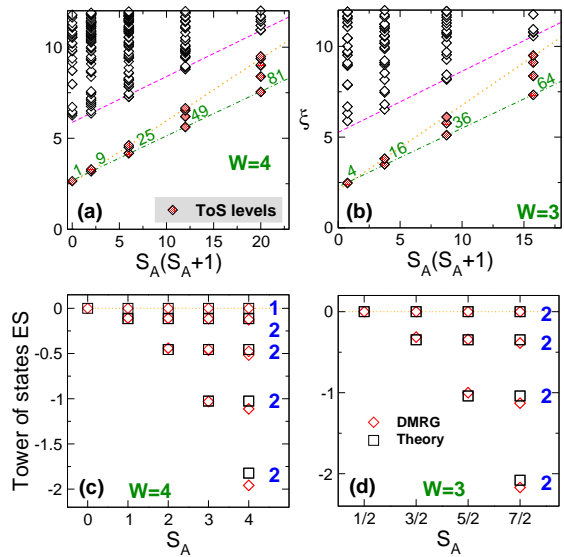}
\caption{ ES of the kagom\'e $J_1$-$J_2$ Heisenberg model (KHA) on 
 the torus. Data is DMRG at $J_2/J_1=-1$ and $W=L/2=3$, $W=L/2=4$ 
 (respectively panels (a) and (b) in the Figure).  The ES is for half of 
 the torus: ES levels $\xi$ plotted versus the subsystem spins $S_A(S_A+1)$. 
 Filled rhombi denote the tower of states (TOS) ES levels. The 
 numbers are the total numbers of TOS levels in each sector $S_A$. 
 The dashed-dotted line highlights the linear behavior (with respect to 
 $S_A(S_A+1)$ of TOS levels). The dashed line marks the higher part of 
 the ES. (c)(d) Enlarged view of TOS structures: TOS ES levels (same data 
 as in (a)(b)) shifted by the value of the highest level (dotted line in 
 (a)(b)) plotted versus $S_A$. The squares denote the (one parameter) 
 fit to the expected result in the large volume limit (cf. 
 formula~\eqref{tos_asy}). The number of degenerate $SU(2)$ multiplets 
 is shown in blue.
}
\label{fig7_torus}
\end{figure}

\subsection{Periodic boundary conditions: ES of the KHA 
on the torus}
\label{boundaries}

Boundary conditions, in particular number of boundaries between the 
two subsystems, can affect dramatically the ES (and the 
entanglement entropies). For instance, in gapped (non-topological) 
one dimensional and two dimensional systems the ES is a boundary 
local quantity~\cite{alba-2011,alba-2012} and a change in the 
 number of boundaries leads to quantitative and qualitative changes 
in the ES. It is interesting to clarify the effect of boundary 
conditions on the TOS structures outlined in~\ref{tos_es}. To this 
purpose here we consider the ES of the $J_1$-$J_2$ KHA on the torus. 

This is illustrated in Figure~\ref{fig7_torus} (ES for half-torus, 
DMRG data at $J_2/J_1=-1$). Data points are for both $W=3$ and $W=4$ (at 
fixed aspect ratio $W/L=1/2$, respectively (a) and (b) in Figure~\ref{fig7_torus}). 
The main features of low-lying ES multiplets are the same as  in the cylindrical 
geometry (compare Figure~\ref{fig7_torus} with Figure~\ref{fig1_tos_ex}).  
The linear behavior of the ES as function of $S_A(S_A+1)$  (Pisa tower 
structure) is clearly visible and an apparent gap divides the low-lying 
ES multiplets from the rest. The number of levels building 
the TOS sector with fixed $S_A$ is given as $(2S_A+1)^2$ (i.e. as for 
kagom\'e cylinders). 

The effective entanglement Hamiltonian ${\mathcal H}_E$ describing the 
TOS structure is given by~\eqref{tos_asy}. This is demonstrated in 
Figure~\ref{fig7_torus} (c,d). ES levels (only TOS levels are shown) 
are plotted versus the block spin $S_A$. Each ES tower (at fixed $S_A$) 
was shifted by subtracting the contribution of the largest level  
(in that sector). Squares are one parameter fits to $\alpha[(S_A^{z'})^2+s_0]$ 
($\alpha$ is the fitting parameter, cf. section~\ref{asy_top}), which give 
$\alpha\approx 0.16$ and $\alpha\approx 0.11$ for respectively $W=3$ and $W=4$. 
It is instructive to observe  that for kagom\'e cylinders one obtains $\alpha
\approx 0.17$ at $W=4$ (cf. section~\ref{asy_top}). The reduction of $\alpha$ 
(by a factor $\approx 2$) has to be attributed to the two boundaries (between 
the subsystems).

\section{Conclusions \& Outlook}
\label{conclusions}

In this Article we studied the {\it ground state} entanglement spectrum 
in $SU(2)$-broken phases. We considered the two dimensional $J_1$-$J_2$ 
Heisenberg model on both the triangular and kagom\'e lattice, restricting 
ourselves to antiferromagnetic(ferromagnetic) $J_1(J_2)$ and $J_2/J_1=-1$.

On both lattices the ground state of the model displays magnetic 
order (and $SU(2)$ symmetry breaking, in the thermodynamic limit). This  
is associated with the appearance in the finite size (spin-resolved) energy 
spectrum of a special type of low-lying excitations, forming the so called 
tower of states (TOS). The TOS structure is divided from the higher part 
of the spectrum (at least for large system sizes) by an energy gap. The number 
of  TOS energy levels in each spin sector $S$ reflects the selected symmetry 
breaking  pattern, and  is given as  $(2S+1)^2$.

In this work we demonstrated that this structure is reflected in the lower  
part of the {\it ground state} ES. Precisely, the ES exhibits families of 
low-lying levels, which are divided from the rest by an {\it entanglement gap}, 
and form a TOS-like structure. The number of TOS levels in a given (subsystem) 
spin sector $S_A$ is $(2S_A+1)^2$, clearly reflecting the corresponding counting 
in the energy TOS. Moreover, finite size behaviors of low-lying ES levels can be 
understood in terms of the energy TOS. All these features can be expressed quantitatively as  
a mapping between the low-lying structure (excitations) of the physical Hamiltonian 
${\mathcal H}$ and of the entanglement Hamiltonian ${\mathcal H}_E$ (expressed by 
formula~\eqref{tos_asy}). 

On the methodological side, our results suggest that entanglement (tower of states) 
spectroscopy, combined with $SU(2)$-symmetric DMRG, could be used as a tool 
for characterizing $SU(2)$-broken phases. Finally, we would like to mention that an 
intriguing research direction originating from this work would be to investigate how 
the TOS structure evolves in the $J_1$-$J_2$ kagom\'e Heisenberg model as the  $J_2=0$ 
point is approached.  In particular, it would be interesting to characterize how the 
low-lying ES levels rearrange to reflect the onset of the $Z_2$ spin liquid 
found in~\cite{white-2011,depenbrock-2012}.

\section{Acknowledgments}

V.A. thanks Andreas L\"auchli for very useful discussions and 
collaboration in a related project. V.A. and S.D. thank the MPIPKS of Dresden,  
where this work was partly done, for the hospitality during the workshop 
``Entanglement Spectra in Complex Quantum Wavefunctions''. U.S. and S.D. 
acknowledge funding by DFG through NIM and SFB/TR 12. 

\end{document}